\documentclass[%
 reprint,
 superscriptaddress,
 amsmath,amssymb,
 aps, physrev,
]{revtex4-2}

\usepackage{graphicx}
\usepackage{dcolumn}
\usepackage{bm}
\usepackage[unicode]{hyperref}

\usepackage[round-mode=figures,round-precision=2,separate-uncertainty=true]{siunitx}

\usepackage{multirow}
\usepackage{cancel}
\makeatletter
\newcommand{\balanceandclearpage}{%
  \par
  \close@column@grid
  \clearpage
  \twocolumngrid
}
\makeatother

 \let\epsilon\varepsilon
 \let\phi\varphi
 \let\theta\vartheta
\def\paramsNrun{32}

\def\paramsimNx{200}
\def\paramsimNy{200}

\def\paramsimdt{\qty{0.002}{}}
\def\paramsimdx{\qty{0.3}{}}

\def\paramsimmodela{\qty{0.8}{}}
\def\paramsimmodelb{\qty{0.05}{}}
\def\paramsimmodeleps{\qty{0.02}{}}

\def\paramsimmodeldiffusivityv{\qty{0.01}{}}

\def\paramnoisetafoursigma{\qty{0.12}{}}

\def\paramnoisetmfoursigma{\qty{0.4}{}}

\def\paramnoisetofoursigma{\qty{0.3}{}}

\def\paramnoisetxafoursigma{\qty{0.75}{}}

\def\paramnoisetxmfoursigma{\qty{1.7}{}}

\def\paramnoisetxofoursigma{\qty{1.0}{}}

\def\paramnoisexafoursigma{\qty{0.1}{}}

\def\paramnoisexmfoursigma{\qty{0.17}{}}

\def\paramnoisexofoursigma{\qty{0.1}{}}

\begin{document}

\preprint{APS/123-QED}

\title{\textbf{Microscopic Variability Alters Macroscopic Rotation Speed in Stochastic Spiral Waves}
}%

\author{Jolien Kamphuis}%
\thanks{These authors contributed equally to this work.}
\affiliation{Mathematical Institute, Leiden University, Leiden, the Netherlands}

\author{Desmond Kabus}%
\thanks{These authors contributed equally to this work.}
\affiliation{Mathematical Institute, Leiden University, Leiden, the Netherlands}
\affiliation{Laboratory of Experimental Cardiology, LUMC, Leiden, the Netherlands}

\author{Hermen Jan Hupkes}%
\email{hhupkes@math.leidenuniv.nl}
\affiliation{Mathematical Institute, Leiden University, Leiden, the Netherlands}

\author{Tim De Coster}%
\email{t.j.c.de\_coster@lumc.nl}
\email{tim.de-coster@ds.mpg.de}
\affiliation{Mathematical Institute, Leiden University, Leiden, the Netherlands}
\affiliation{Laboratory of Experimental Cardiology, LUMC, Leiden, the Netherlands}
\affiliation{Max Planck Institute for Dynamics and Self-Organisation, G\"ottingen, Germany}

\date{\today}

\begin{abstract}
We present a general theory for noise-induced corrections to the angular velocity of spiral waves. Stochasticity produces two second-order effects: an instantaneous term from heterogeneity that always slows rotation, and an orbital-drift term from temporal fluctuations that can either accelerate or decelerate it. For our parameters, orbital drift is weaker, producing a net slowdown. Analytical predictions match Barkley-model simulations with temporal noise. Examination of additional noise types \textit{in silico} confirms angular velocity slowing. This mechanism provides a robust route by which stochasticity reshapes spiral dynamics in excitable media, with direct implications for arrhythmias and neural wave propagation.
\end{abstract}

\maketitle


\section{\label{sec:introduction}Introduction}

\noindent Spiral waves are a fundamental phenomenon observed in various physical, chemical, and biological systems \cite{murray2007mathematical}. These self-sustaining, rotating wave patterns emerge in excitable media, playing a critical role in diverse disciplines, from cardiac electrophysiology \cite{panfilov2019spiral} to reaction-diffusion chemistry \cite{winfree1972spiral}. Despite their idealized representations in mathematical models, real-world spiral waves are seldom perfect. Instead, they exhibit irregularities due to various forms of noise that permeate natural and experimental systems \cite{tsimring2014noise}, e.g. thermal fluctuations.

Noise in spiral wave systems can manifest in multiple ways, including temporal noise, spatial noise, and spatio-temporal noise \cite{alonso2001noise}. Temporal noise refers to fluctuations in time, disrupting the periodicity of wave dynamics. Spatial noise introduces heterogeneities across the medium, altering the structural coherence of the spiral. Spatio-temporal noise, a combination of the two, affects both the propagation and stability of the wave, leading to complex dynamical behaviors. These perturbations influence system evolution in non-trivial ways and may have far-reaching consequences for both theoretical understanding and practical applications \cite{song2023impact, steinbock1993control}.

In this work, we investigate the impact of noise on spiral waves. By introducing controlled perturbations, we analyze how noise modifies the core properties of these waves, leading to shifts in their trajectories and changes in their rotation speed. We employ complementary analytical and numerical approaches: the analytical framework decomposes the noise-induced correction into instantaneous and orbital-drift contributions, while \textit{in silico} simulations of the Barkley model provide direct verification and mutual validation. Together with recent stability results \cite{kuehn2022stochastic}, these methods yield a coherent picture of how stochasticity reshapes spiral-wave dynamics in excitable media.

\section{\label{sec:results}Methods and Results}

\noindent We first derive a theoretical prediction for arbitrary reaction–diffusion systems and noise statistics [Eq.~(\ref{eq:pred_omega_sigma})]. For validation, we picked the Barkley model subjected to temporal white noise, which allowed for a comparison between the numerical evaluation of the analytical result [Eq.~(\ref{eq:numerical})] and direct computations [Eq.~(\ref{eq:computational})].

\begin{figure*}
\includegraphics[width=\textwidth]{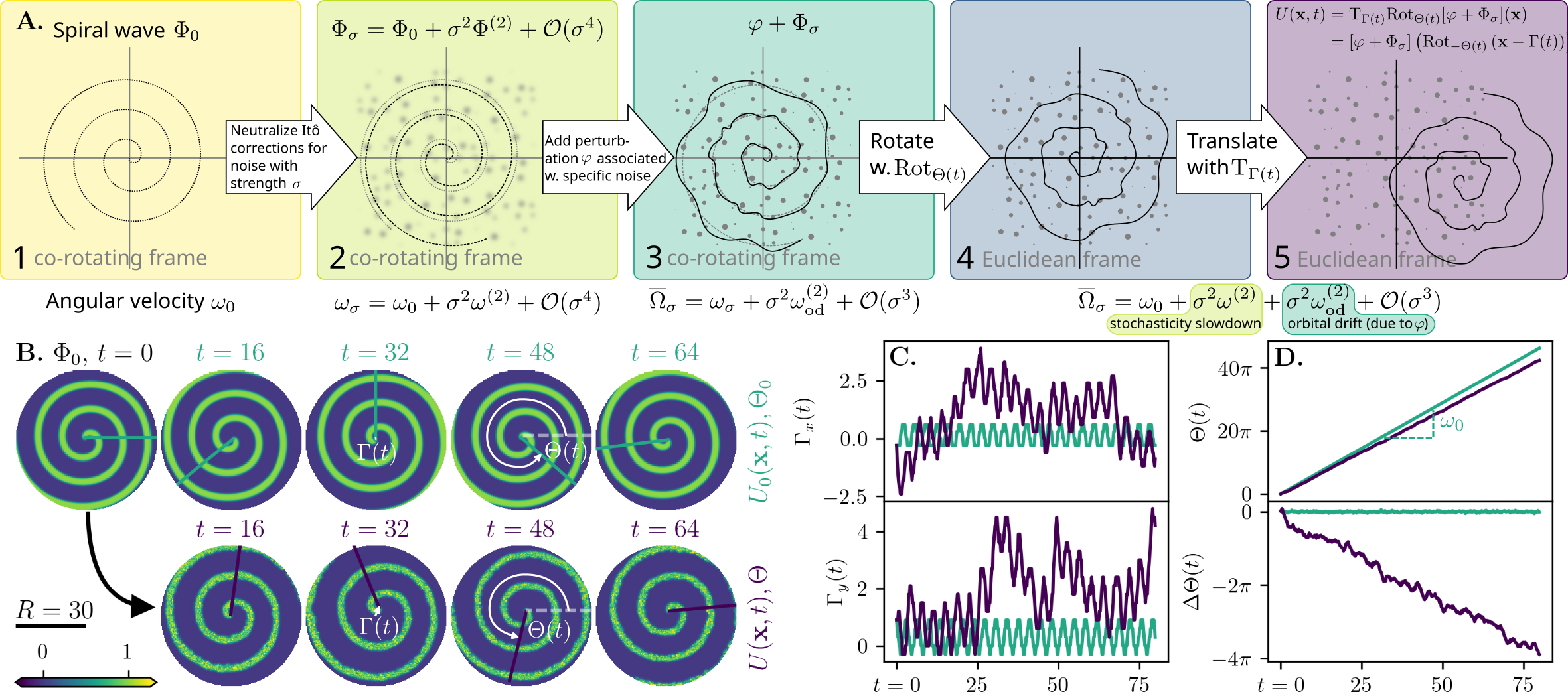}
\caption{\label{fig:mathscheme}
        \textbf{Decomposition and dynamics of stochastic spiral waves.}
        (\textbf{A}) Schematic visual aid to help with the five-step decomposition \eqref{eq:decomp:U} of $U$ involving the coordinates $\Theta$ and $\Gamma$ \eqref{eq:ortho:cond} and the instantaneous stochastic spiral $(\omega_\sigma,\Phi_\sigma)$.
        (\textbf{B}) Initial condition and numerical time evolution of deterministic and stochastic (multiplicative spatio-temporal noise, $\sigma=1.7$) spirals \cite{barkley1990spiral}.
        (\textbf{C}) Temporal evolution of spiral coordinates.
        (\textbf{D}) Rotation angle $\Theta(t) = \Omega(t)\cdot t$ over time for deterministic and stochastic spirals; lower panel shows the deviation from a linear fit to the deterministic spiral $\Delta \Theta(t) = \Theta(t) - \omega_0 \cdot t$.
}
\end{figure*}

\subsection{Stochastic Spiral Waves}\label{sec:stochasticspiralwaves}
\noindent One of the methods to describe spiral waves in excitable and oscillatory media is making use of reaction-diffusion equations (RDEs).
\paragraph{Stochastic Reaction-Diffusion Equation:}
We consider a stochastic partial differential equation (SPDE), more specifically a two-dimensional RDE given by \cite{Hamster_Hupkes_2020, van2025multidimensional, bosch2025local}:
\begin{equation}
    \mathrm{d}U = \left[\rho \nabla^{2}U + f(U)\right]\mathrm{d}t + \sigma h(U) \mathrm{d}W^{Q}(t), \label{eq:readiff}
\end{equation}
with $U(\textbf{x}, t) \in \mathbb{R}^{n}$, $\textbf{x} \in \mathbb{R}^{2}$, $\rho \in \mathbb{R}^{n \times n}$ a diagonal matrix with non-negative elements, $f: \mathbb{R}^{n} \rightarrow \mathbb{R}^{n}$ describing local dynamics, $\sigma \in \mathbb{R}_{\geq 0}$ a scalar determining the noise strength, $h: \mathbb{R}^{n} \rightarrow \mathbb{R}^{n}$ a noise function, and $W^{Q}(t)$ a general (cylindrical)-Q-Wiener process \cite{lototsky2017stochastic, karczewska2005stochastic, da2014stochastic}. When $\sigma~=~0$, we assume that the resulting system admits an asymptotic spatially-periodic structure as the solution to Eq.~(\ref{eq:readiff}), i.e. a deterministic spiral wave $\Phi_0$ with a constant rotation speed $\omega_{0}$ \cite{sandstede2023spiral} that we write as
\begin{equation}
\label{eq:det:spiral}
U(x, t) = \mathrm{Rot}_{\omega_0 t} [\Phi_0](x)    
\end{equation}
When $\sigma$ is small, the stochastic spiral wave solution $U$ stays close to the deterministic spiral wave. However, when $\sigma$ is large, spiral breakup could occur or the spiral wave could vanish.

For our theoretical predictions we assume that the deterministic spiral \eqref{eq:det:spiral} is spectrally stable: the only spectrum in the right-half plane consists of eigenvalues $0$ and $\pm i\omega$ associated to the translational and rotational symmetry of the system \cite{dodson2019determining}, with adjoint eigenfunctions $\psi_{0}$ and $\psi_{\pm}$ that are exponentially localized \cite{sandstede2023spiral}. In this setting spiral waves are very robust against perturbations, as long as these are centered away from the core. In particular, the perturbed dynamics can be well-understood by projecting onto the three-dimensional eigenspace associated to the rotational and translational symmetries \cite{sandstede1997dynamics}, leading to shifts in the phase and position of the spiral.
Computationally, we confirm that our qualitative predictions extend into parameter regimes where spiral meandering occurs as the consequence of a Hopf bifurcation, in which case the spectral stability is lost.

\paragraph{Decomposition of $U$:}
We compare the SPDE solution $U(\mathbf{x},t)$ and its angular velocity $\Omega(t)$ with the deterministic rotating wave solution $\Phi_{0}$ and $\omega_{0}$ (Fig.~\ref{fig:mathscheme}A shows the solution in a co-rotating frame at angular velocity $\omega_{0}$).

To this end, we introduce a stochastic reference profile $\Phi_{\sigma}$ with angular velocity $\omega_{\sigma}$ in the co-rotating frame (Fig.~\ref{fig:mathscheme}A-2). To describe stochastic corrections, we introduce $\Omega(t) = \Theta(t) / t \in\mathbb{R}$ as the stochastic angular velocity with $\Theta(t)\in \mathbb{R}$ the stochastic angle (Fig.~\ref{fig:mathscheme}A-4) and $\Gamma(t) = (\Gamma_{x}, \Gamma_{y})^{T}\in\mathbb{R}^{2}$ as a stochastic translation (Fig.~\ref{fig:mathscheme}A-5). We then define a perturbation $\varphi$ in the co-rotating frame (Fig.~\ref{fig:mathscheme}A-3) such that
\begin{eqnarray}
\label{eq:decomp:U}
    U(\mathbf{x},t) &=& \mathrm{T}_{\Gamma(t)} \,\mathrm{Rot}_{\Theta(t)} \big[ \varphi + \Phi_{\sigma} \big](\mathbf{x}) \nonumber \\
    &=& \big[\varphi + \Phi_{\sigma}\big]\!\left(\mathrm{Rot}_{-\Theta(t)}(\mathbf{x}-\Gamma(t))\right),
\end{eqnarray}
referring to IIAc below for the precise definitions.
Transforming this equation shows that $\varphi(t)$ denotes the deviation of $U$ from $\Phi_{\sigma}$ after translating $U$ by $-\Gamma(t)$ and rotating by $-\Theta(t)$, capturing the effect of temporal noise fluctuations beyond the instantaneous profile ($\Phi_{\sigma}$, $\omega_{\sigma}$).

\paragraph{Essential definitions and properties:}
In order to assign a precise meaning to the 
spiral rotation angle $\Theta$ and displacement $\Gamma$,
we impose the orthogonality conditions
\begin{equation}
\label{eq:ortho:cond}
    \langle \varphi, \psi_0 \rangle = \langle \varphi, \psi_+ \rangle = \langle \varphi, \psi_- \rangle = 0
\end{equation} 
between the spiral perturbation and the adjoint eigenfunctions.
As long as the perturbation $\varphi$ is small, this uniquely fixes all the terms in the decomposition \eqref{eq:decomp:U} apart from $\Phi_\sigma$. By applying the It\^o lemma \cite{da2019mild, morters2010brownian}
we naturally uncover the dynamics (see Appendix~\ref{appendix:math})
\begin{equation}
    \left( \begin{array}{c} d\Theta \\ d \Gamma \end{array}\right) = \mathcal{C}_\sigma(U, \Theta, \Gamma) \, dt + \sigma \mathcal{R}(U, \Theta, \Gamma) d W_t^Q,
\label{eq:defns:eq:om:gamma:cal:p:cal:q}
\end{equation}
where $\mathcal{C}$ and $\mathcal{R}$ contain the deterministic and stochastic terms respectively. In addition, the behaviour of the perturbation $\varphi$ can be described by
\begin{equation}
\mathrm{d}\varphi = \mathcal{D}_{\sigma}(\varphi)\mathrm{d}t + \sigma\mathcal{S}_{\sigma}(\varphi)\mathrm{d}\widetilde{W}^{Q}_{t}\label{eq:dphi},
\end{equation}
where $\mathcal{D}_{\sigma}(\varphi)$ (deterministic contributions) and $\mathcal{S}_{\sigma}(\varphi)$ (stochastic contributions) are nonlinear functions of $\varphi$ and its derivatives (see Eq.~(\ref{eq:def:cal:r:s})) and $\widetilde{W}_t^Q$ is an adjusted version of the noise process that takes into account rotations and translations, but has the same statistical behavior as $W_t^Q$. The term $\mathcal{D}_\sigma$ implicitly depends on $(\omega_\sigma, \Phi_\sigma)$, which can now be fixed by imposing $\mathcal{D}_\sigma(0) = 0$, neutralizing the second order It{\^o} corrections.
In view of Eq.~(\ref{eq:dphi}), this means that the perturbation $\varphi$ stemming from the choice $U(0) = \Phi_\sigma$ will initially only feel stochastic forcing. This also holds for the deviation $\Theta(t) - \omega_\sigma t$, which is why we refer to the pair $( \Phi_\sigma, ~\omega_\sigma)$ as the \textit{instantaneous} stochastic wave.

\paragraph{Instantaneous expansion:}
We will first look at the correction terms that are required to obtain the stochastic wave $\Phi_{\sigma}$ with angular velocity $\omega_{\sigma}$ from the deterministic wave $\Phi_{0}$ with angular velocity $\omega_{0}$. By expanding Eq.~(\ref{eq:def:cal:r:s}) through substitution of third-order Taylor expansions of $\Phi_{\sigma}$ and $\omega_{\sigma}$ around $\sigma$, it can be seen that first and third-order terms are zero. We hence obtain:
\begin{eqnarray}
    \Phi_{\sigma} &&= \Phi_{0} + \cancel{\sigma\Phi^{(1)}} + \sigma^{2}\Phi^{(2)} + \cancel{\sigma^{3}\Phi^{(3)}} + \mathcal{O}(\sigma^{4})\label{eq:phi_sigma_expansion},\\
    \omega_{\sigma} &&= \omega_{0} + \cancel{\sigma\omega^{(1)}} + \sigma^{2}\omega^{(2)} + \cancel{\sigma^{3}\omega^{(3)}} + \mathcal{O}(\sigma^{4})\label{eq:omega_sigma_expansion}.
\end{eqnarray}
By inspecting the first component of $\mathcal{C}_\sigma$, one can extract bilinear maps $\mathcal{B}_1$ and $\mathcal{B}_2$ (see Eq.~(\ref{eq:defs:b:1:b:od})) such that
\begin{equation}
\begin{array}{lcl}
    \omega^{(2)} & = & \mathrm{Tr} \, \mathcal{B}_1[ h(\Phi_0), \mathcal{R}(\Phi_0,0,0)]
    \\[0.2cm]
    & & \quad
    + \mathrm{Tr} \, \mathcal{B}_2[ \mathcal{R}(\Phi_0,0,0), \mathcal{R}(\Phi_0,0,0)],
\end{array}
\label{eq:id:for:omega:0:2}
\end{equation}
thereby showing that the second order correction to the instantaneous angular velocity has its origin in the initial stochastic contributions to the spiral rotation and displacement. Here the $Q$-trace operator should be interpreted as
\begin{equation}
\mathrm{Tr} \, \mathcal{B}_1[ h,q]
= \sum_{k=0}^\infty \mathcal{B}_1[ h \sqrt{Q} e_k, q \sqrt{Q} e_k]
\end{equation}
and similarly for $\mathcal{B}_2$, where
$(e_k)_{k \ge 0}$ denotes an (arbitrary) orthonormal basis of the Hilbert space from which the noise is sampled (see Appendix \ref{appendix:noise}).

\paragraph{Orbital drift expansion:}
The next step is to examine the fluctuations of our quantities of interest around the instantaneous stochastic spiral wave $(\Phi_{\sigma}, \omega_{\sigma})$, which we refer to as ‘orbital drift’ \cite{hamster2020stability}.
Inspecting the definitions of $\mathcal{D}_\sigma$ and $\mathcal{C}_\sigma$, our co-rotating reference frame enables us to take $t \to \infty$ and introduce the
limiting quantities (see \cite{bosch2025conditional} for rigorous interpretations in 1D)
\begin{equation}
\begin{aligned}
    \varphi_{\sigma, \rm od} &= \lim_{t \to \infty} \mathbb{E}(\text{Rot}_{\Theta(t)}\text{T}_{-\Gamma(t)}U(t)) -\Phi_{\sigma},\\
    \omega_{\sigma, \rm od} &= \lim_{t \to \infty} \mathbb E (t^{-1} \Theta(t) )-\omega_{\sigma},
    \label{eq:orbitaldrift}
\end{aligned}
\end{equation}
where the expectation $\mathbb{E}$ averages over fluctuations. Analogous limits for $\Gamma$ are not possible, since the second and third components of $\mathcal{C}_\sigma$ converge to a limiting quantity that rotates with $\Omega(t)$. As above, we are interested in the expansions
\begin{equation}
\begin{aligned}
    \varphi_{\sigma, \rm od} &&= \cancel{\sigma\varphi_{\rm od}^{(1)}} + \sigma^{2}\varphi_{\rm od}^{(2)} + \cancel{\sigma^{3}\varphi_{\rm od}^{(3)}} + \mathcal{O}(\sigma^{4}),
    \\
    \omega_{\sigma, \rm od} &&= \cancel{\sigma\omega_{\rm od}^{(1)}} + \sigma^{2}\omega_{\rm od}^{(2)} + \cancel{\sigma^{3}\omega_{\rm od}^{(3)}} + \mathcal{O}(\sigma^{4})\label{eq:orb_drift_expansion}.
\end{aligned}
\end{equation}
where the odd terms vanish because the system remains invariant upon taking $\sigma \to - \sigma$ and $W^Q \to - W^Q$.
In order to compute the coefficients in Eq.~(\ref{eq:orb_drift_expansion}), we need to expand the dynamics of the perturbation $\varphi$ in the form
\begin{equation}
    \varphi(t) = \sigma \varphi^{(1)}(t) + \sigma^{2} \varphi^{(2)}(t) + \mathcal{O}(\sigma^{3}).\label{eq:varphi}
\end{equation}
Extracting the linear part of $\mathcal{D}_\sigma$ by decomposing Eq.~(\ref{eq:def:cal:r:s}) as
\begin{equation}
    \mathcal{D}_\sigma(\varphi) =
    \underbrace{ \rho \nabla^2 \varphi
    + \omega_0 \partial_\theta \varphi +  Df(\Phi_0) \varphi}_{L [\varphi]}
     + \mathcal{O}\big( \| \varphi \|^2 + \sigma^2 \| \varphi \| \big),
\end{equation}
and isolating the $O(\sigma)$ terms in Eq.~(\ref{eq:dphi}) after substituting Eq.~(\ref{eq:varphi}) in it, we arrive at
\begin{equation}
    d \varphi^{(1)} = L [\varphi^{(1)}] \, dt
     + \mathcal{S}_0(0) \, d \widetilde{W}^Q_t.
    \label{eq:varphi:1:spde}
\end{equation}
Writing $(e^{L s})_{s \ge 0}$ for the semigroup generated by $L$, the solution to Eq.~(\ref{eq:varphi:1:spde}) can be explicitly written in the mild form
\begin{equation}
    \varphi^{(1)}(t) = \int_0^t e^{L s} \mathcal{S}_0(0) d \widetilde{W}_s^Q.
\end{equation}
Expanding the phase difference $\Delta \Theta(t) = \Theta(t) - \omega_0 t$ and the position $\Gamma(t)$ in a similar fashion,
the leading $O(\sigma)$-contributions satisfy
\begin{equation}
    \left( \begin{array}{c} d \Theta^{(1)} \\ d \Gamma^{(1)} \end{array}\right) =
    \mathcal{K}(\omega_0 t, 0)[\Phi_0]^{-1} \Pi(0,0) h(\Phi_0)
    d \widetilde{W}^Q_t .
\end{equation}
This implies that all three components are (scaled) Brownian motions,
reminiscent of the chaotic deterministic behaviour uncovered in hypermeandering regimes \cite{biktashev1998deterministic, ashwin2001hypermeander}.
In particular, $\mathbb E \big(\Theta^{(1)}(t) , \Gamma_x^{(1)}(t), \Gamma_y^{(1)}(t) \big) = (0, 0, 0)$ and one may also compute $\mathbb E \big( \varphi^{(1)}(t) \big) = 0$.

This changes at second order in $\sigma$, since quadratic expressions involving $\varphi^{(1)}$ typically do not have vanishing expectations. Indeed, by inspecting the $O(\sigma^2)$ terms in the first component of $\mathcal{C}_\sigma$, one can extract a bilinear map $\mathcal{B}_{\rm od}$ (see Eq.~(\ref{eq:defs:b:1:b:od})) so that
\begin{equation}
   \omega_{\rm od}^{(2)} = \lim_{t \to \infty} \mathbb{E} \Big( \mathcal{B}_{\rm od}[\varphi^{(1)}(t), \varphi^{(1)}(t) ] \Big) .
\end{equation}
Using the mild It{\^o} formula \cite{da2019mild}, this can be evaluated as
\begin{equation}
    \omega_{\rm od}^{(2)} =
     \int_0^\infty \mathrm{Tr} \, \mathcal{B}_{\rm od} [ e^{Ls} \mathcal{S}_0(0), e^{Ls} \mathcal{S}_0(0) ] \, ds .
\label{eq:comp:od:with:semi:grp}
\end{equation}
In contrast with the instantaneous angular velocity, the second order correction to the orbital drift rotation speed has its origin in the initial stochastic contribution to the spiral perturbation.

In the end we are interested in the total average angular velocity $\overline{\Omega}_\sigma$ of a stochastic spiral wave, which we obtain by transforming Eq.~(\ref{eq:orbitaldrift}) into
\begin{eqnarray}
    \overline{\Omega}_\sigma &&= \omega_\sigma + \sigma^2\omega_{\sigma, \rm od}^{(2)}+\mathcal{O}(\sigma^3)\nonumber\\
    &&= \omega_0 + \sigma^2 [\omega^{(2)} + \omega^{(2)}_{\rm od}] +\mathcal{O}(\sigma^3)\label{eq:pred_omega_sigma}.
\end{eqnarray}
Note that this prediction is valid regardless of the noise type or the specific reaction diffusion model chosen.

\paragraph{Meandering spiral waves}
In the meandering regime where spectral stability is lost, the integral in \eqref{eq:comp:od:with:semi:grp} will typically diverge due to the new non-decaying modes in the semigroup $(e^{Ls})_{s \ge 0}$ associated with the eigenvalues that cross at the Hopf-bifurcation. To correct for this, one can introduce two new variables $R_m$ and $\Theta_{m}$ that account for the amplitude and phase of the meandering mode and study the five-component analogue of Eq.~\eqref{eq:defns:eq:om:gamma:cal:p:cal:q}.

\subsection{Validation}

\noindent We are able to validate our theoretical predictions by picking a specific reaction-diffusion model and a specific type of noise. Because of its general applicability, we chose the Barkley model \cite{barkley1990spiral}, which we modified to include noise terms:
\begin{eqnarray}\label{eq:Barkley}
    \mathrm{d} u  = &&\left[ ~\nabla^2 u + \tfrac{1}{\epsilon} u(1-u)\left(u-\tfrac{v+b}{a}\right) \right] \mathrm{d} t \nonumber\\
           &&+ \sigma (h_0+h_uu) \xi(t)\sqrt{\mathrm{d}t},\\
    \mathrm{d} v  = &&\left[ \delta \nabla^2 v + u-v\right] \mathrm{d}t,
\end{eqnarray}
with parameters $a=\paramsimmodela$, $b=\paramsimmodelb$, $\epsilon=\paramsimmodeleps$ and $\delta=\paramsimmodeldiffusivityv$. The Barkley model uses arbitrary units in time and space. Noise-less simulations with $\xi = 0$ result in $\omega_{0} = 1.819$. The scalar parameters $h_0$ and $h_u$ control the dependence of the noise on the state variable $u$. We can use them to create additive, multiplicative, and attenuating noise (also known as negative multiplicative noise, or stochastic damping noise) in Eq.~(\ref{eq:Barkley}):
\begin{itemize}
    \item $h_0 = 1$ and $h_u = 0$: additive noise,
    \item $h_0 = 0$ and $h_u = 1$: multiplicative noise,
    \item $h_0 = 1 = - h_u$: attenuating noise.
\end{itemize}
The choice of $h_{0}$ and $h_{u}$ allows to numerically \cite{shardlow2005numerical} compute (Appendix~\ref{appendix:comp1}) Eq.~(\ref{eq:pred_omega_sigma}) on large bounded disks with homogeneous Neumann boundary conditions. For zero-mean unit-variance normal-distributed multiplicative temporal noise this results in:
\begin{equation}\label{eq:numerical}
\begin{aligned}
    \overline{\Omega}_\sigma &= \omega_{0} + \sigma^{2}[\omega^{(2)}+\omega^{(2)}_{\rm od}]+\mathcal{O}(\sigma^3)\\
    &= 1.82 + \sigma^{2}[-0.90 - 0.10] +\mathcal{O}(\sigma^3)\\
    &= 1.82 - 1.00\sigma^{2}.
\end{aligned}
\end{equation}
Additive and attenuating temporal noise give values of -11.77 and -5.89 for $\omega^{(2)}$, and 0.69 and 0.61 for $\omega^{(2)}_{\rm od}$ respectively. Hence, while the instantaneous term $\omega^{(2)}$ always causes rotational speed slowdown, the orbital drift term $\omega^{(2)}_{\rm od}$ can both accelerate or slow down rotation.

\begin{figure}
\includegraphics[width=\columnwidth]{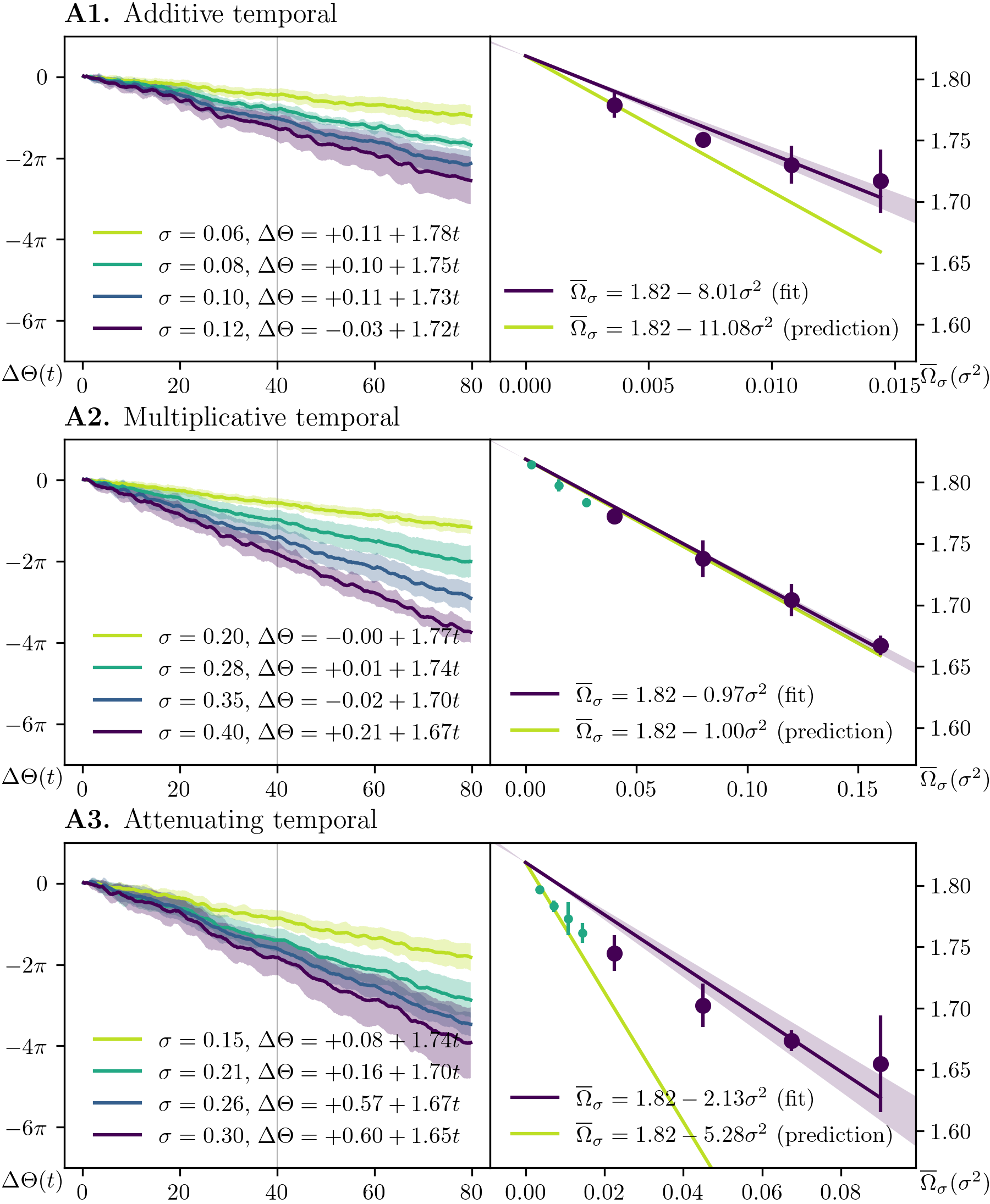}
\caption{\label{fig:multipltempnoise}
        \textbf{Effects of temporal noise on the angular velocity of spiral waves in the Barkley model.}
        (\textbf{left}) Average of the angular difference $\Delta\Theta$ from the noiseless reference for four noise levels $\sigma$ (\paramsNrun{} runs each); shaded regions show one standard deviation. Linear fits exclude transients.
        (\textbf{right}) Fitted angular velocities $\overline{\Omega}_\sigma$ demonstrate spiral slowing, consistent with theoretical predictions.
}
\end{figure}

To validate this calculation, we assessed stochastic spirals from a computational modeling standpoint (Appendix~\ref{appendix:comp2}). By simulating a variety of spiral waves with different noise levels (Fig.~\ref{fig:mathscheme}B), we can track their movements in time (Fig.~\ref{fig:mathscheme}C). Splitting up this movement in translations (Fig.~\ref{fig:mathscheme}D) and rotations (Fig.~\ref{fig:mathscheme}E), we can look at the angular difference with a deterministic reference spiral without noise. For every noise level, we fitted the average of 32 simulations to obtain an angular velocity (Fig.~\ref{fig:multipltempnoise}A).
These angular velocities allowed us to find a relation between the stochastic angular velocity and the deterministic one, depending on the noise level (Fig.~\ref{fig:multipltempnoise}A2):
\begin{equation}\label{eq:computational}
    \overline{\Omega}_\sigma = 1.82 - 0.97\sigma^{2}
\end{equation}
The two other kinds of temporal noise also agree closely with our mathematical derivation when we consider small values of $\sigma$ (Fig.~\ref{fig:multipltempnoise}A3, where extra measurements were taken (green) for attenuating noise in order to stay in the linear regime), confirming the validity of the theoretical framework. Deviations from a straight line at higher $\sigma$ are likely due to higher order noise effects $\mathcal{O}(\sigma^{3})$. Hence we have shown by two independent approaches that noise can slow spiral rotation, with microscopic fluctuations encoded in $\varphi$ driving orbital drift. It should be noted that while the orbital drift term is not large enough to overcome the second order instantaneous angular velocity correction, orbital drift contributions can either cooperate or counteract the instantaneous correction term.

\subsection{Generalisation}
\label{subsec:generalisation}

\paragraph{Noise types:}
Using computational modeling we investigated different types of noise besides temporal for the term $\xi$ (see Appendix~\ref{appendix:noise}), namely spatial, and spatio-temporal white noise (zero-mean unit-variance normal-distributed). For each of these noise types, we once again made a distinction between three kinds of noise: additive, multiplicative, and attenuating noise.

The fitted second-order correction terms $\omega^{(2)}+\omega_{\rm od}^{(2)}$ are: temporal (-8.01, -0.97, -2.13), spatial (-16.13, -2.22, -19.79), and spatio-temporal (-0.11, -0.05, -0.01) for additive, multiplicative, and attenuating noise, respectively. All second order correction terms in $\sigma$ have opposite sign to the angular velocity $\omega_{0} = 1.819$. Therefore all different noise types and kinds show varying levels of slowdown with increasing noise level (Fig.~\ref{fig:allnoise}). Out of the different noise types, spatial noise slows down most, and spatio-temporal least. When looking at the different kinds of noise, it is additive noise that slows down the most, and multiplicative the least.

\begin{figure}
\includegraphics[width=\columnwidth]{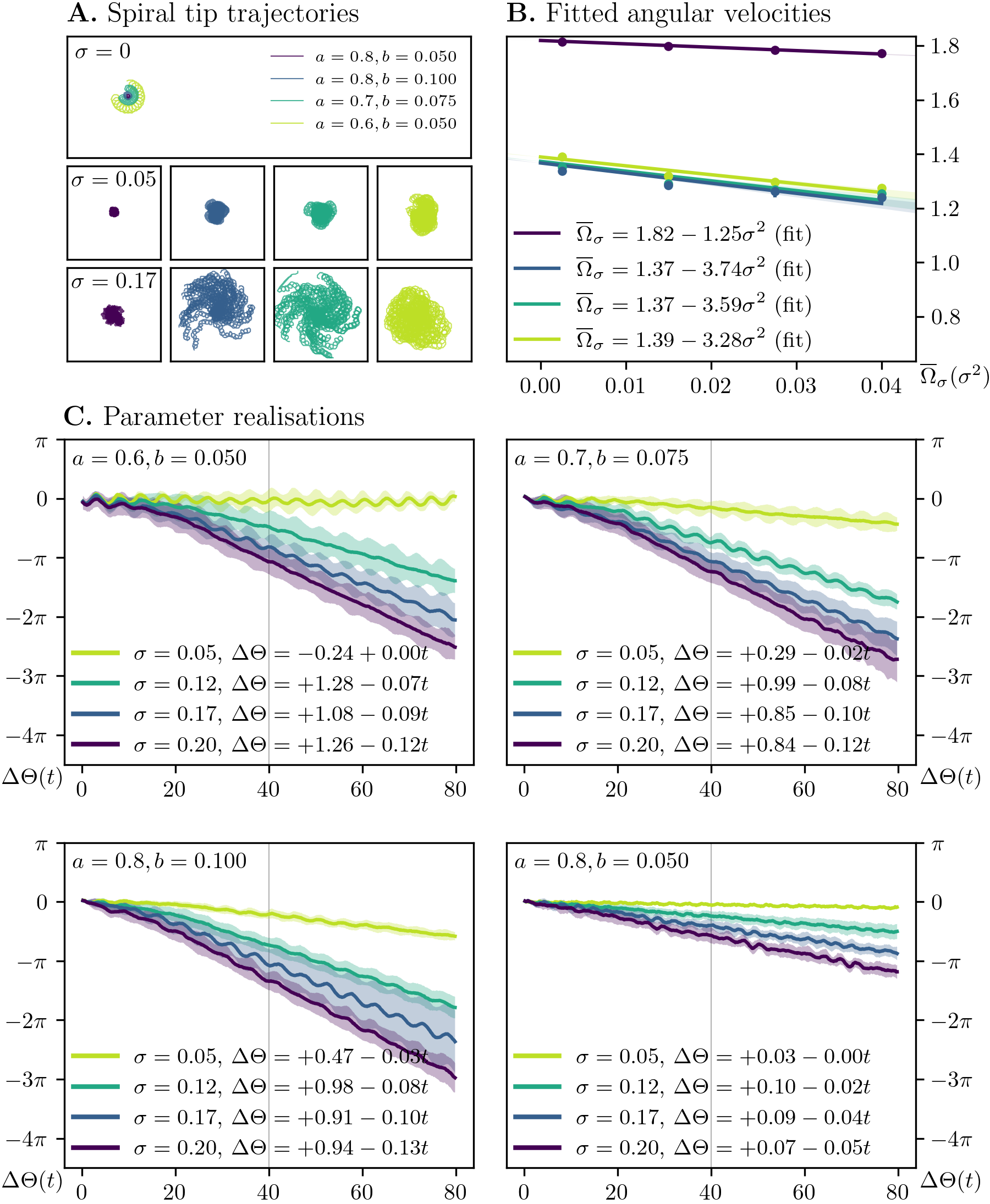}
\caption{\label{fig:tm.meandering}
        \textbf{Effects of multiplicative temporal noise on the angular velocity of meandering spiral waves in the Barkley model.}
        (\textbf{A}) Phase singularity tracking of spiral waves.
        (\textbf{B}) Fitted angular velocities $\overline{\Omega}_\sigma$ demonstrate more severe spiral slowing for meandering spirals.
        (\textbf{C}) Average of the angular difference $\Delta\Theta$ from the noiseless reference for four noise levels $\sigma$ (\paramsNrun{} runs each) under different parameters; shaded regions show one standard deviation. Linear fits exclude transients.
}
\end{figure}

\paragraph{Meandering:}
Besides looking at spirals under different kinds of noise, different parameters for the Barkley model were also tried for multiplicative temporal noise (Fig.~\ref{fig:tm.meandering}C), especially within the meandering regime outlined in \cite{luo2024unpinning} and including the parameters shown in Fig.~\ref{fig:multipltempnoise}A2. Making these spirals stochastic resulted in hypermeandering consistent with \cite{biktashev1998deterministic} (Fig.~\ref{fig:tm.meandering}A), and once again a decrease in angular velocity (Fig.~\ref{fig:tm.meandering}B). Note that the decrease is stronger for meandering spirals, but nearly independent of the chosen parameters. This stronger decrease hints at a correlation between hypermeandering (displacement) and additional angular velocity slowdown, most likely attributable to a second orbital drift term associated to the new phase.

\section{\label{sec:discussion}Discussion}

\noindent We have demonstrated that stochastic spiral waves undergo rotation speed altering irrespective of the noise type, where orbital drift can both accelerate and decelerate the spiral wave rotation.

Excitable media such as chemical, biological, and physical systems share the property that spiral and scroll waves govern their macroscopic dynamics \cite{murray2007mathematical}. Cardiac tissue is a prominent example of such a system, where spiral waves of electrical activity underlie tachyarrhythmias and fibrillation \cite{gray1998spatial,gray1995mechanisms}. In the following we focus on cardiology as a representative application field. We find that microscopic fluctuations can have a macroscopic influence on the angular velocity of spiral waves, revealing a double-edged character of stochastic slowing. On one hand, reduced rotation rates can be beneficial, since slower spirals are less prone to break up into life-threatening fibrillatory turbulence \cite{ten2003influence, qu1999cardiac, fenton2002multiple}. On the other hand, slower spirals are more likely to anchor to tissue heterogeneities, making them more resistant to termination \cite{davidenko1992stationary, defauw2014small, takagi2004unpinning}.

The influence of noise depends sensitively on the Barkley-model parameters. For the parameters used here, multiplicative noise mainly perturbs the waveback, attenuating noise the wavefront, and additive noise both regions. Because the wavefront is more gradual, its response dominates. These differences may explain the opposite sign of the orbital-drift terms. Although the instantaneous term was only observed to dominate the orbital drift term, parameter regimes might be found where the orbital-drift contribution exceeds the instantaneous term, producing a net acceleration of spiral rotation. Perturbations of the wavefront and waveback are known to alter spiral meandering, analogous to changes in ionic currents such as increased sodium or calcium conductance or reduced potassium conductance \cite{qu2000origins}. Stochastic modulation of local excitability may therefore distort the instantaneous rotation orbit, contributing to irregular meandering observed experimentally.

Our results further show that the influence of noise depends on its type: spatiotemporal noise is the most robust, allowing large amplitudes before propagation failure or spiral breakup occurs, while purely spatial or temporal noise destabilizes waves at much lower noise levels (Fig.~\ref{fig:allnoise}). However, noise has a larger influence on the slowdown of purely spatial or temporal noise than it has on spatiotemporal noise. This distinction has direct implications for cardiac tissue, where natural spatiotemporal fluctuations may shift toward predominantly spatial noise with aging or fibrosis \cite{dhein2021remodeling, panfilov2002spiral}. In such cases, spirals may initially slow further, but when the noise level rises too high, spiral breakup becomes inevitable. Importantly, static spatial heterogeneity is already known to cause conduction slowing \cite{nezlobinsky2020anisotropic, de2018arrhythmogenicity} and break-up, albeit before a spiral has formed. Once a spiral has formed, our findings suggest that in addition to this baseline effect, stochastic perturbations can further stabilize and slow spiral waves.

Lastly, we investigated the influence of the noise amplitude $\sigma$ on the observed rotational speed $\Omega$, given the ``true angular velocity" $\omega_{0}$. This relation is valuable for predicting experimental noise effects from simulation data. Conversely, by measuring $\Omega$ at different noise levels and performing a linear fit in $\sigma^{2}$, extrapolation to $\sigma=0$ yields predictions of $\omega_{0}$ with an accuracy of $\approx 98\%$ (fits not shown). This approach thus enables inference of the ``true rotational speed" from noisy biological samples and could be incorporated into uncertainty quantification frameworks for cardiac tissue dynamics.

Beyond cardiology, our results may also be relevant for other excitable media, such as cortical tissue in neuroscience \cite{roxin2005role}, chemical oscillatory systems \cite{winfree1972spiral}, and bacterial or gut microbial waves \cite{tse2016electrophysiological}, where noise and heterogeneity likewise play central roles in wave dynamics.

\begin{acknowledgments}
\noindent This work was supported by the Netherlands Organisation for Scientific Research (NWO Open Mind grant 2025/TTW/02025375 to TDC, and Vici 10.61686/NMHBE05119 to HJH). Additional funding came from the European Union's Horizon Europe research and innovation programme (under the Marie Skłodowska-Curie grant agreement No 101205806 to TDC).
\end{acknowledgments}

\section*{Data Availability}
\noindent The data that support the findings of this article are openly available \cite{CodeStochasticSpirals}.

\bibliographystyle{apsrev4-1}
\bibliography{main}

\balanceandclearpage
\appendix
\label{appendix}

\section{Mathematical definitions}
\label{appendix:math}

Here we summarize the computations in \cite{kamphuis2022stochastic} to provide explicit definitions for the functions appearing in the main text.
The index $i$ always ranges through the set $\{0, -, +\}$,
while the indexes $j$ and $k$ are always taken from $\{ \theta, x, y \}$, implicitly summing over duplicates.
We first introduce the $\mathbb{R}^3$-valued
functions $\Pi$ and $\mathcal{G}$ that act as
\begin{equation}
\begin{aligned}
\relax [\Pi(\Theta, \Gamma) w]_i & = \langle w, T_\Gamma \mathrm{Rot}_\Theta \psi_i \rangle ,\\
\mathcal{G}(\Theta, \Gamma)[u, q, \tilde{q}]_i & = \tfrac{1}{2} \langle u, T_{\Gamma} \partial_{jk} \mathrm{Rot}_\Theta  \psi_i \rangle q_j \tilde{q}_k,
\end{aligned}
\end{equation}
together with the $\mathbb{R}^{3 \times 3}$-valued
function $\mathcal{K}$ that acts as
\begin{equation}
    \mathcal{K}(\Theta, \Gamma)[u]_{i,j} = \langle u, T_\Gamma \partial_j \mathrm{Rot}_\Theta \psi_i \rangle.
\end{equation}
The function $\Pi$ encodes projections with respect to the adjoint eigenfunctions, while $\mathcal{K}$ and $\mathcal{G}$ can be seen as the first and second derivatives of these projections with respect to the rotation $\Theta$ and translation $\Gamma$.
We observe that the defining orthogonality conditions $\langle \varphi, \psi_0 \rangle = \langle \varphi, \psi_+ \rangle = \langle \varphi, \psi_- \rangle = 0$ can be written as
\begin{equation}
    \Pi(\Theta, \Gamma) U  = \Pi(\Theta, \Gamma) T_\Gamma \mathrm{Rot}_\Theta \Phi_\sigma.
\end{equation}
Noting that the right-hand side is constant in time and applying the It\^o lemma to the left-hand side, the stochastic term $\mathcal{R} = (\mathcal{R}_{\theta}, \mathcal{R}_x, \mathcal{R}_y)^T$ in Eq.~(\ref{eq:defns:eq:om:gamma:cal:p:cal:q}) can be written as
\begin{equation}
    \mathcal{R}(U, \Theta, \Gamma) =  \mathcal{K}(\Theta,\Gamma)[U]^{-1} \Pi(\Theta, \Gamma) h(U).
\end{equation}
This allows the deterministic term to be expressed as
\begin{align}
    \mathcal{C}_\sigma&(U, \Theta, \Gamma) = \nonumber\\
    & \mathcal{K}(\Theta, \Gamma) [U]^{-1} \Pi(\Theta, \Gamma) [ \rho \nabla^2 U + f(U)] \\
    & - \sigma^2 \mathrm{Tr} \, \mathcal{K}(\Theta, \Gamma) [U]^{-1}
    \mathcal{K}(\Theta, \Gamma) [h(U)][\mathcal{R}(U, \Theta,\Gamma) ] \nonumber\\
    & + \sigma^2 \mathrm{Tr} \, \mathcal{K}(\Theta, \Gamma) [U]^{-1} \mathcal{G}(\Theta, \Gamma)[U, \mathcal{R}(U, \Theta,\Gamma),\mathcal{R}(U, \Theta,\Gamma)]. \nonumber
\end{align}
In particular, the bilinear forms used to determine $\omega^{(2)}$ and $\omega_{\rm od}^{(2)}$ in Eq.~(\ref{eq:id:for:omega:0:2}) and Eq.~(\ref{eq:comp:od:with:semi:grp}) can be represented explicitly as
\begin{equation}
\begin{aligned}
\mathcal{B}_1[w, q] &= -e_0^T \mathcal{K}(0,0)[\Phi_0]^{-1} \mathcal{K}(0,0)[w][q],\\
\mathcal{B}_2[q, \tilde{q}] &= e_0^T
\mathcal{K}(0, 0) [\Phi_0]^{-1}
\mathcal{G}(0, 0)[\Phi_0, q, \tilde{q}],\\
\mathcal{B}_{\rm od}[ \varphi, \tilde{\varphi}]
&= \tfrac{1}{2} e_0^T  \mathcal{K}(0,0)[\Phi_0]^{-1} \Pi(0,0) f''(\Phi_0)[ \varphi, \tilde{\varphi}],
\end{aligned}
\label{eq:defs:b:1:b:od}
\end{equation}
in which $e_0^T = (1, 0, 0)$ extracts the first component of 3-vectors.
Applying the It\^o lemma to the representation
\begin{equation}
    \varphi = \mathrm{Rot}_{-\Theta} T_{-\Gamma} U - \Phi_\sigma
\end{equation}
and introducing the shorthand $\mathbb U = \Phi_\sigma + \varphi$, we can finally write the nonlinearities
in Eq.~(\ref{eq:dphi}) in the form
\begin{equation}
\begin{aligned}
    \mathcal{D}_\sigma( \varphi)  =
    &f(\mathbb U) + \rho \nabla^2 \mathbb U +  \mathcal{C}_{\sigma,j}(\mathbb U, 0, 0) \partial_j \mathbb U\\
& + \sigma^2 \mathrm{Tr} \,  \mathcal{R}_j(\mathbb U, 0, 0) \partial_j h(\mathbb U)\\
& + \tfrac{1}{2} \sigma^2
   \mathrm{Tr} \, \mathcal{R}_j( \mathbb U, 0, 0) \mathcal{R}_k( \mathbb U, 0, 0) \partial_j \partial_k \mathbb U,\\
\mathcal{S}_\sigma( \varphi) =
& h(\mathbb U) + \mathcal{R}_j(\mathbb U, 0, 0) \partial_j \mathbb U .
\end{aligned}
\label{eq:def:cal:r:s}
\end{equation}

\section{Noise realization}
\label{appendix:noise}
For temporal noise, the function $\xi$ in Eq.~(\ref{eq:Barkley}) is sampled from a zero-mean unit-variance normal distribution at each timestep. In this case $Q$ and $\mathrm{Tr}$ both reduce to identity operators. For spatial-temporal noise, one chooses an orthonormal basis $(e_k)_{k \ge 0}$ for the space of square-integrable functions, picks a spatial covariance function $q$ and writes $Qv = q*v$ for the associated convolution operator.
For white noise one substitutes the delta function for $q$. In Eq.~(\ref{eq:Barkley}) one uses the sum $\xi =  \sqrt{Q} e_k \xi_k$ where each $\xi_k$ is again sampled from a zero-mean unit-variance normal distribution at each time. Spatial noise is sampled in the same fashion at $t = 0$ but remains constant thereafter, while $\sqrt{dt}$ is replaced by $dt$ in Eq.~(\ref{eq:Barkley}).

\begin{figure*}
\includegraphics[width=\textwidth]{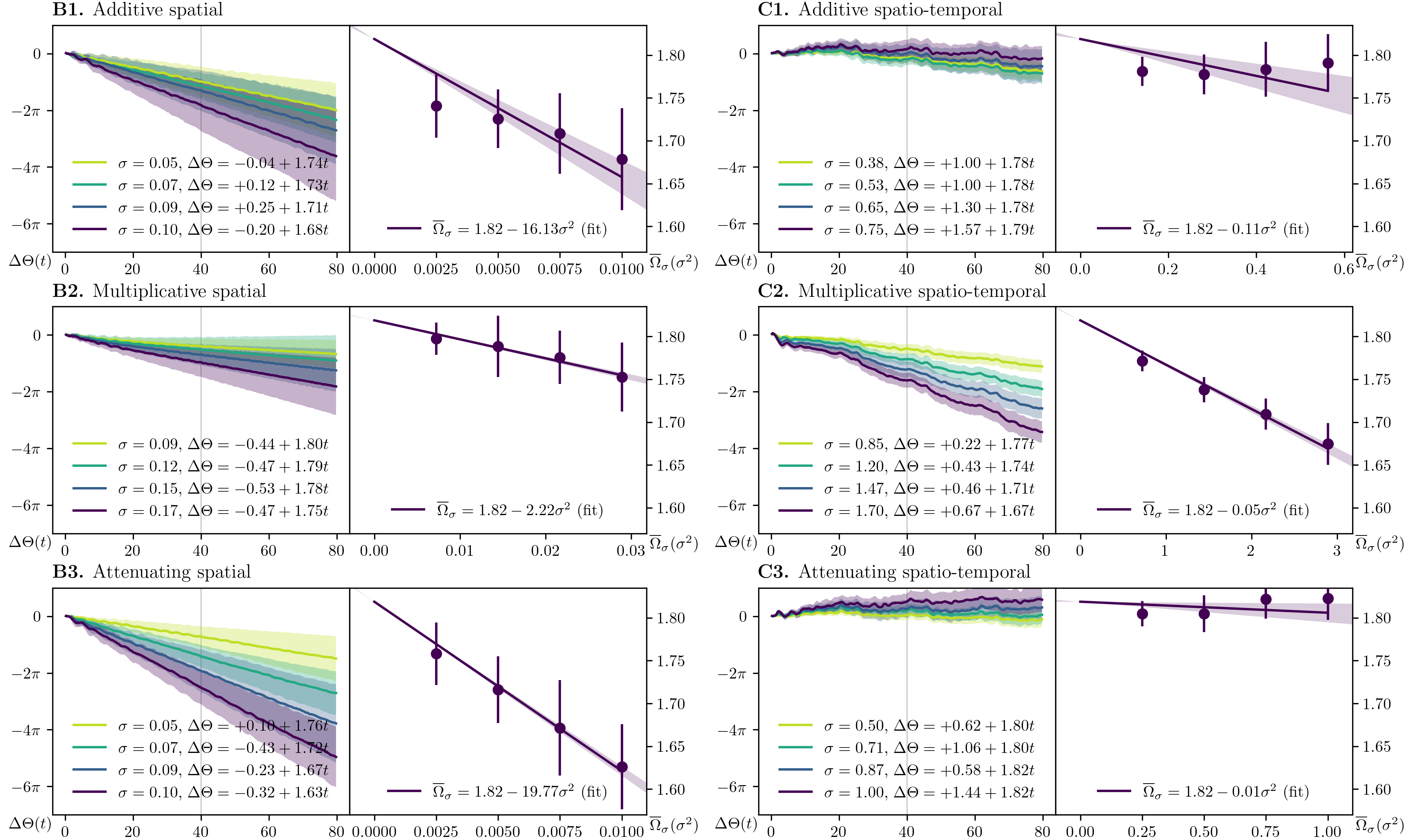}
\caption{\label{fig:allnoise}
        \textbf{
            Effects of spatio(-temporal) noise on the angular velocity of spiral waves in the Barkley model.
        }
        We consider four different noise levels $\sigma$ with \paramsNrun{} finite-differences simulations each.
        The average of the angular difference $\Delta\Theta$ to the reference is plotted for each $\sigma$.
        The shaded area corresponds to one standard deviation around the mean.
        Linear functions are fitted to the second half of the temporal evolution to exclude transient effects (left).
        Linear functions are fitted to the data to obtain expressions $\overline{\Omega}_\sigma$ (right). Also spatial (\textbf{B}) and spatio-temporal (\textbf{C}) noise are observed to slow down a spiral wave.
}
\end{figure*}

\section{Coefficient computation}
\label{appendix:comp1}
The expressions for $\omega^{(2)}$ and $\omega^{(2)}_{\rm od}$ were evaluated by considering a circular domain with radius $R = 50$. The adjoint eigenfunctions $(\psi_0, \psi_-, \psi_+)$ were computed using a $192\times 100$ polar $(\theta, r)$ discretization. The semigroup evaluation
in Eq.~(\ref{eq:comp:od:with:semi:grp}) was performed on this grid using the trapezium method with $dt = 0.001$, projecting out the non-decaying modes at each timestep. The inner products and derivatives appearing in the bilinear forms Eq.~(\ref{eq:defs:b:1:b:od}) were evaluated on a $1800\times1800$ Carthesian grid.

\section{Spiral simulations}
\label{appendix:comp2}
\noindent We numerically solve the SPDE of Eqs.~(\ref{eq:Barkley}) with initial and Neumann boundary conditions using finite-differences. For this, we use Pigreads, the Python-integrated GPU-enabled reaction-diffusion solver \cite{kabus2026pigreads}. We use a forward Euler time stepping scheme with a step of $\Delta t = \paramsimdt$. In space $\vec x = [x, y]^\text{T}$, we use a square grid with $\paramsimNx\times\paramsimNy$ points with a grid spacing of $\Delta x = \Delta y = \paramsimdx$. The Laplacian is approximated using a five-point stencil taking the Neumann boundary conditions for a disk of radius $R=30$ into account. The simulations were run on an NVIDIA GeForce RTX 2080 SUPER Mobile / Max-Q GPU.

Noise is implemented according to Eq.~\ref{eq:Barkley} by adding stochastic terms
$h_0$ and $h_u$, which are independent Gaussian random variables with zero mean
and unit variance, i.e.\ $h_0, h_u \sim \mathcal{N}(0,1)$. Random numbers are
generated using a linear congruential generator combined with the Box-Muller
transform in polar form \cite{teukolsky1992numerical}.

We perform an initial run without noise ($\xi = 0$, $\sigma = 0$, or equivalently $h_0 = 0 = h_u$) inducing a spiral wave which is approximately centered in the domain. All other runs start from the last frame of this initial run. For each of the nine types of noise, we consider four different noise levels $\sigma$, with \paramsNrun{} simulations each. The maximum value of $\sigma$ for each type of noise was chosen as the highest value where the single spiral wave remains stable, i.e., where it does not break up into multiple waves and does not disappear through interaction with the boundary. Lower noise levels $\sigma$ are chosen evenly spaced in $\sigma^2$ between 0 and the maximum value. For temporal noise, the maximum noise levels are \paramnoisetafoursigma{}, \paramnoisetmfoursigma{}, and \paramnoisetofoursigma{} for additive, multiplicative, and attenuating noise, respectively. For spatial, they are \paramnoisexafoursigma{}, \paramnoisexmfoursigma{}, and \paramnoisexofoursigma{}; and for spatio-temporal \paramnoisetxafoursigma{}, \paramnoisetxmfoursigma{}, and \paramnoisetxofoursigma{}, respectively.

Subsequently, we localize the spiral wave tip using phase defect detection \cite{kabus2022numerical} as a numerical proxy for the orthogonality conditions in Eq.~(\ref{eq:ortho:cond}). For more robustness to noise, we blur $V$ and $R$ in space with a Gaussian filter of standard deviation $\sigma = 2$~px. We then use the cosine method to calculate the phase defect density $\rho(\vec x)$ \cite{kabus2022numerical,tomii2021spatial}. To further increase the robustness of the tip localization, we multiply $\rho(\vec x)$ with a Gaussian bell function, centered at the previous tip location $\vec x_{\text{tip}}$. We use the grid point with highest $\rho'(\vec x)$ as the new tip location. We call the evolution of the tip location over time the spiral trajectory.

Let $\vec y = \vec x - \vec x_{\text{tip}} = [x - x_{\text{tip}}, y - y_{\text{tip}}]^\text{T}$ be the spatial coordinates relative to the tip. We determine the rotation angle $\Theta(t)$ around the tip over time by minimizing the norm of the difference between the values on the circle $C$ with a radius of 20 pixels around the tip in the last frame of the reference run -- which is the zeroth frame of all other runs $u(0, \vec x)$ -- and the current frame $u(t, \vec x)$ rotated by a candidate angle $\theta'$. As the studied spiral wave rotates counter-clockwise, i.e. $\omega > 0$, we can restrict the search for the rotation angle to the interval $\mathcal{I}(t) = \Theta(t - \Delta t) + [0, \pi/4]$:
\begin{equation}
    \Theta(t)
    =
    \arg\min_{\theta' \in \mathcal{I}(t)}
    \left\lVert
        u(0, \vec y) - \mathrm{Rot}_{\theta'}(u(t, \vec y))
    \right\rVert_{\vec y \in C},
\end{equation}
in which the rotation is numerically implemented with nearest-neighbor interpolation.

Combining all \paramsNrun{} simulations in each noise level, we calculate the average of the rotation angle $\Theta(t)$ over time. We fit a linear function to the second half of its temporal evolution to exclude transient effects when changing from the noise-free initial condition to the noise-perturbed spiral wave. The slope of this linear function is the average angular velocity $\overline{\Omega}_\sigma$ of the spiral wave (Fig.~\ref{fig:allnoise}).

Finally, for each type of noise, we fit a linear function to the average angular velocity $\overline{\Omega}_\sigma$ over the noise level $\sigma^2$, forcing the curve through $\omega_{0}$ and taking the uncertainty of the angular velocity into account (Fig.~\ref{fig:allnoise}).

\end{document}